\pdfoutput=1

\documentclass[sigconf,natbib=true,anonymous=false]{acmart}

\setcopyright{none}
\renewcommand\footnotetextcopyrightpermission[1]{}

\usepackage[T1]{fontenc}

\usepackage[hang,flushmargin]{footmisc}

\usepackage{url}            
\usepackage{booktabs}       
\usepackage{amsfonts}       
\usepackage{nicefrac}       
\usepackage{microtype}      
\usepackage{amsmath}
\usepackage{enumitem}
\usepackage{graphicx}       
\usepackage{caption}
\usepackage{subcaption}
\usepackage{threeparttable}
\usepackage{color, colortbl}
\usepackage{threeparttable,booktabs}
\usepackage{multirow}
\usepackage{balance}

\newcommand\ignore[1]{}

\newcommand{\mytt}[1]{\texttt{\small #1}}

\title{Resources for Brewing BEIR: Reproducible Reference Models\\ and an Official Leaderboard}

\author{Ehsan Kamalloo,$^1$ Nandan Thakur,$^1$ Carlos Lassance,$^2$ \\
Xueguang Ma,$^1$ Jheng-Hong Yang,$^1$ Jimmy Lin$^1$}
\affiliation{\vspace{0.1cm}$^1$ David R. Cheriton School of Computer Science, 
University of Waterloo \country{Canada}\\
$^2$ Naver Labs Europe \country{France}}

\begin{document}

\renewcommand{\shortauthors}{}
\pagestyle{empty}

\begin{abstract}
BEIR is a benchmark dataset for zero-shot evaluation of information retrieval models across 18 different domain/task combinations.
In recent years, we have witnessed the growing popularity of a representation learning approach to building retrieval models, typically using pretrained transformers in a supervised setting.
This naturally begs the question:\ How effective are these models when presented with queries and documents that differ from the training data?
Examples include searching in different domains (e.g., medical or legal text) and with different types of queries (e.g., keywords vs.\ well-formed questions).
While BEIR was designed to answer these questions, our work addresses two shortcomings that prevent the benchmark from achieving its full potential:\
First, the sophistication of modern neural methods and the complexity of current software infrastructure create barriers to entry for newcomers.
To this end, we provide reproducible reference implementations that cover the two main classes of approaches:\ learned dense and sparse models.
Second, there does not exist a single authoritative nexus for reporting the effectiveness of different models on BEIR, which has led to difficulty in comparing different methods.
To remedy this, we present an official self-service BEIR leaderboard that provides fair and consistent comparisons of retrieval models.
By addressing both shortcomings, our work facilitates future explorations in a range of interesting research questions that BEIR enables.
\end{abstract}

\maketitle

\section{Introduction}

One recent conceptual innovation in information retrieval is the recognition that the ``classic'' task of {\it ad hoc} retrieval can be framed as a representation learning problem.
\citet{dpr} demonstrated that transformers can be used to encode queries and documents into a dense vector space, where top-$k$ retrieval translates into the problem of nearest neighbor search.
This led to the development of many so-called dense retrieval models~\cite{dpr,ance,Hofstatter_etal_SIGIR2021,colbert,rocketqa,rocketqav2}.
Separately, \citet{Zamani_etal_CIKM2018} showed that neural networks can be used to learn sparse representations of queries and documents that are amenable to retrieval using standard inverted indexes.
Later, researchers applied transformers for learning these sparse lexical representations, which led to a long line of so-called sparse retrieval models~\cite{sparterm,jang-etal-2021-ultra,unicoil,splade,spade,spladev2,slim}.

\begin{figure}[t]
\centering
\includegraphics[width=0.4\textwidth]{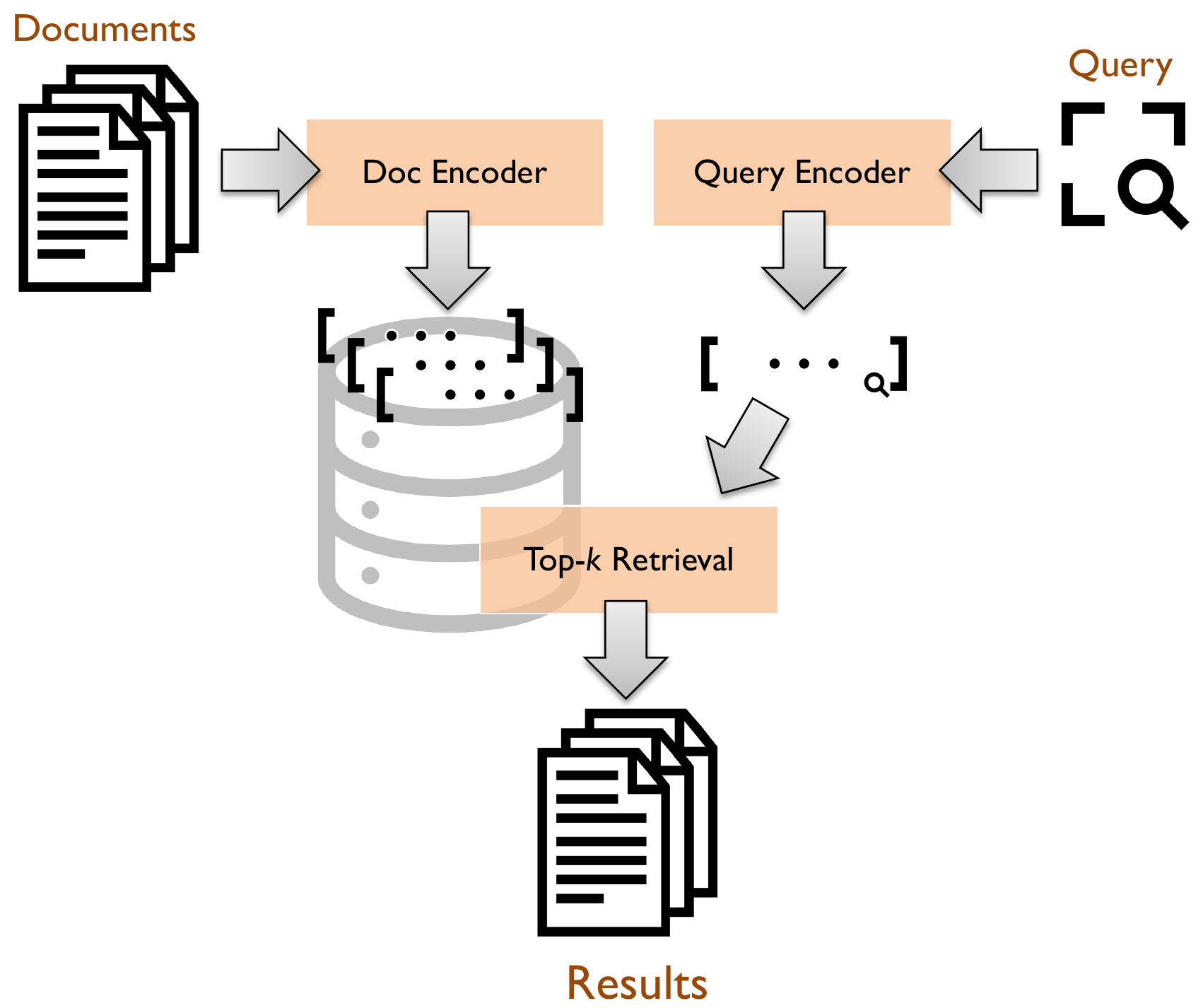}
\caption{A bi-encoder architecture for information retrieval that captures both dense retrieval models (e.g., DPR), sparse retrieval models (e.g., SPLADE), as well as traditional ``bag-of-words'' lexical retrieval models (e.g., BM25).}
\label{fig:bi-encoders}
\end{figure}

\citet{Lin_SIGIRForum2021} recently pointed out that learned dense representations, learned sparse representations, and even traditional lexical retrieval models such as BM25 can be viewed as parametric variations of a bi-encoder architecture (see Figure~\ref{fig:bi-encoders}).
In this design, both queries and documents are fed to ``encoders'' that generate vector representations.
Retrieval boils down to the problem of efficiently finding the top-$k$ most similar document representations given a query representation and a similarity function, usually the inner product.

The design of encoders in such a bi-encoder architecture is dictated primarily by two choices:\ (1) the basis of the vector space and (2) how the vector weights are assigned.
For example, both dense models such as DPR and sparse models such as SPLADE use pretrained transformers to encode queries and documents into vectors; both take advantage of large amounts of manually labeled data.
However, the critical difference is the representational basis of their vectors---DPR generates dense vectors, typically with the same width as the output contextual embeddings (for many models, 768 dimensions), whereas SPLADE ``projects'' the scalar weights of each dimension back into the input vocabulary space, generating, in essence, bag-of-words vectors.
BM25 can be understood in this bi-encoder architecture as having a document ``encoder'' that was heuristically designed (i.e., the BM25 scoring function) and a query ``encoder'' that generates multi-hot vectors.

Viewing retrieval as representation learning not only helps us understand the relationship between different models, but immediately illuminates open research questions.
The dominant approach today is based on supervised learning with (manually) labeled datasets such as the MS MARCO test collections~\cite{msmarco}.
This naturally begs the question:\ What happens when models are applied to out-of-distribution data?
Examples include applying retrieval models trained on one type of text (e.g., passages from the web) to another type of text (e.g., text from the medical or legal domain), or differences between training and test queries (e.g., keyword queries vs.\ well-formed natural language questions).

This is where BEIR ({\it Be}nchmarking
{\it IR})~\cite{beir} comes in.
BEIR is a benchmark for zero-shot evaluation of information retrieval models that enables exactly the types of explorations outlined above.
For example, BEIR has shown that traditional lexical retrieval models such as BM25 remain competitive baselines---and in fact, the first dense retrieval models evaluated on BEIR were worse overall than BM25 in a zero-shot setting.
With BEIR, researchers have discovered that sparse retrieval models appear to achieve better cross-domain generalization than dense retrieval models.
The dataset has spurred entirely new lines of research---for example, on {\it unsupervised} representation learning~\cite{Izacard:2112.09118:2021}, where BEIR served as the benchmark for demonstrating model effectiveness.

Nevertheless, we can identify two shortcomings in the state of the current BEIR ecosystem.
First, the variety and complexity of modern neural retrieval models create barriers to entry for researchers who wish to explore the research questions that BEIR enables.
Successfully executing an end-to-end retrieval run requires coordinating heterogeneous software components that differ depending on the model type.
It would be desirable to have reproducible implementations of retrieval models that are easily accessible to everyone, particularly newcomers.

Second, there does not exist an authoritative nexus for sharing and comparing the effectiveness of models on BEIR.
When the benchmark was first introduced, the creators shared a publicly readable spreadsheet that tabulated results from different models.
Since there was no mechanism for updating this spreadsheet beyond personal communications, it soon fell into disuse.

Today, the {\it de facto} method for disseminating and comparing BEIR results is via tables in research papers, with copied-and-pasted effectiveness figures from previous work.
This, needless to say, is a laborious and error-prone process.
Over time, issues emerged, the most concerning of which is the reporting of results on partially overlapping subsets of the 18 datasets that comprise BEIR.
This practice erodes the benefits of BEIR---the entire point of the benchmark is to cover the widest possible range of domains---and makes comparisons of different models difficult.
It would be desirable if there existed a single, ``official'' nexus that gathered all BEIR evaluation results from the community for comparison purposes.
In other words, it would be nice to have a self-service central leaderboard.

\paragraph{Contributions}
This work builds on BEIR and aims to address the two main shortcomings discussed above.
We make the following contributions:

\begin{itemize}[leftmargin=*]

\item We share with the community reproducible implementations of five popular retrieval models for BEIR in the open-source Pyserini IR toolkit.
From our extensive documentation pages, an end-to-end retrieval run can be reproduced with only two clicks:\ copy and paste of a command-line invocation.

\item We describe the methodological innovation of using radar charts to visualize the effectiveness of different retrieval models across the BEIR datasets.
These visualizations allow a researcher to quickly pinpoint the source of gains and losses with respect to a baseline, providing an entry point for error analyses.

\item We present an official self-service leaderboard for BEIR that unifies community-wide evaluation efforts.
Our leaderboard is built on the EvalAI platform and provides an authoritative nexus for sharing results in a consistent manner.

\item We explore novel variations of existing retrieval models that examine field indexing, wordpiece tokenization, sliding window techniques for handling long documents, and hybrid fusion.
Analyses of these variants with radar charts provide additional insights into model effectiveness.

\end{itemize}

\begin{figure}[t]
\centering
\includegraphics[width=0.4\textwidth]{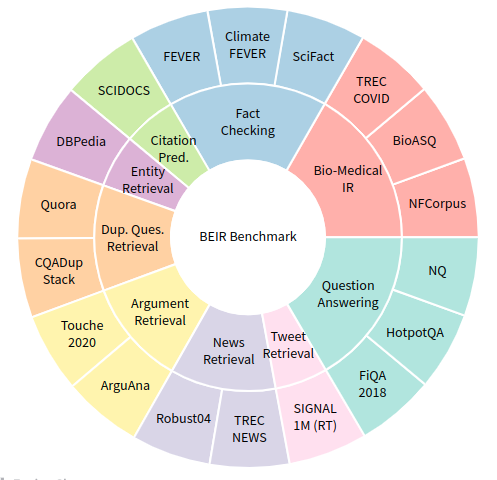}
\caption{Overview of the 18 datasets that comprise the BEIR benchmark.}
\label{fig:beir-overview}
\end{figure}

\section{BEIR Overview}

The BEIR benchmark, introduced by~\citet{beir}, is designed to evaluate information retrieval systems across diverse combinations of tasks and domains.
It was designed to focus on the ``zero-shot'' retrieval setting, i.e., evaluation on tasks and domains without any training data or supervision signals.
The benchmark enables researchers to explore the out-of-domain generalization capabilities of retrieval models and has steered innovation toward more robust and adaptable retrieval methods.

An overview of the tasks and domains covered by BEIR across 18 distinct datasets is shown in Figure \ref{fig:beir-overview}.
In terms of task diversity, BEIR spans the gamut from traditional {\it ad hoc} retrieval tasks (e.g., the TREC 2004 Robust Track) to, for example, Natural Questions (NQ) \cite{nq}, an open-domain question answering task involving retrieval of English Wikipedia passages that can answer natural language questions.
BEIR tasks further include argument retrieval (e.g., Argu\-Ana, T\'ouche-2020) and fact checking (e.g., FEVER, SciFact), both of which are related to, but distinct from, traditional {\it ad hoc} retrieval.
BEIR also covers various domains, including scientific articles, news, Wikipedia, tweets, and so on.
Finally, queries take different forms and can vary in length, from a few keywords \cite{boteva2016} to as long as a paragraph \cite{wachsmuth:2018a}. 

Table~\ref{tab:beir_stats} summarizes the 18 datasets that comprise BEIR.
The datasets range in the amount of relevance judgments available.
A few datasets have ``dense'' judgments, such as TREC-COVID \cite{10.1093/jamia/ocaa091}, with 66k judgments for 50 test queries, but many have ``sparse'' judgments, such as SciFact, with only 339 judgments.
The corpora associated with the datasets also vary in size, some containing millions of passages, whereas others have only a few thousand.
BEIR also standardizes its evaluation metric and uses nDCG@10 and recall@100 across all datasets to compare the effectiveness of each system on an equal footing.
Individual dataset scores are macro-averaged across all datasets for a final cumulative score. 

\begin{table}[t]
\centering
    \small
    \resizebox{0.48\textwidth}{!}{\begin{tabular}{ l | r | r | r | c | c}
        \toprule
           \textbf{Dataset} &\textbf{\#Q} & \textbf{\#J} & \textbf{\#Passages} & \textbf{Task} & \textbf{Domain} \\
         \midrule
         \textbf{TREC-COVID} &  50   & 66,336 & 171,332    & \multirow{3}{*}{Bio-Medical IR} & \multirow{3}{*}{Bio-Medical} \\
         \textbf{BioASQ}     & 500   & 2,359 & 14,914,602 &  &  \\
         \textbf{NFCorpus}   & 323   & 12,334 & 3,633      &  &  \\ \midrule
         \textbf{NQ}         & 3,452 & 4,201 &  2,681,468 & \multirow{3}{*}{QA} & Wikipedia\\
         \textbf{HotpotQA}   & 7,405 & 14,810 &  5,233,329 & & Wikipedia\\
         \textbf{FiQA-2018}  & 648   & 1,706 &  57,638 & & Finance \\ \midrule
         \textbf{Signal-1M (RT)} & 97 & 1,899 & 2,866,316 & Tweet-Retrieval & Twitter \\ \midrule
         \textbf{TREC-NEWS}      & 57 & 15,655 & 594,977   & \multirow{2}{*}{News-Retrieval} & \multirow{2}{*}{News} \\
         \textbf{Robust04}       & 249 & 311,410 & 528,155 & & \\ \midrule
         \textbf{ArguAna}        & 1,406 & 1,406 & 8,674  & \multirow{2}{*}{Argument-Retrieval} & \multirow{2}{*}{Misc.} \\
         \textbf{Touch\'e-2020}  & 49 & 2,214 &   382,545  & & \\ \midrule
         \textbf{CQADupStack}    & 13,145 & 23,703 &  457,199   & \multirow{2}{*}{Dup. Ques.-Retrieval} & StackExc. \\
         \textbf{Quora}    & 10,000 & 15,675 & 522,931  & & Quora \\ \midrule
         \textbf{DBPedia}  & 400    & 43,515 & 4,635,922 & Entity-Retrieval & Wikipedia \\ \midrule
         \textbf{SCIDOCS}  & 1,000 & 29,928 & 25,657 & Citation-Prediction & Scientific \\ \midrule
         \textbf{FEVER}    & 6,666 & 7,937 &  5,416,568 & \multirow{3}{*}{Fact Checking} & Wikipedia \\
         \textbf{Climate-FEVER}    & 4,681 & 4,682 &  5,416,593 &  & Wikipedia \\
         \textbf{SciFact}    & 300 & 339 &  5,183 &  & Scientific \\
    \bottomrule
    \end{tabular}}
    \vspace{0.25cm}
    \caption{Summary of the 18 datasets that comprise the BEIR benchmark. \#Q and \#J denote the total counts of queries and relevance judgments in the test split of each dataset.}
    \label{tab:beir_stats}
\end{table}

The BEIR authors provided additional resources with the benchmark.
They shared a GitHub repository\footnote{\url{https://github.com/beir-cellar/beir}} that contains source code for the evaluation framework along with example usage.
The code is written in Python and is available on PyPI (\texttt{pip install beir}).
With the original release of BEIR, the authors hosted a leaderboard in the form of a Google spreadsheet.\footnote{\url{https://docs.google.com/spreadsheets/d/1L8aACyPaXrL8iEelJLGqlMqXKPX2oSP_R10pZoy77Ns/edit}}
There were many issues with this that led to the replacement we describe in Section~\ref{section:leaderboard}.

\section{Retrieval Models}
\label{section:models}

This work provides reproducible reference implementations of five different retrieval models for BEIR.
These comprise a ``bag-of-words'' BM25 baseline, two learned dense retrieval models (TAS-B and Contriever), and two learned sparse retrieval models (uniCOIL without expansion and SPLADE).
In this section, we provide an overview of these models as they are presented in the literature, but explore different model variants in Section~\ref{section:variants}.

\subsection{Multi-Field BM25}

Despite tremendous progress in neural retrieval, ranking using traditional lexical ``bag-of-words'' models such as BM25 remains a strong baseline.

The original BEIR paper presented a BM25 baseline using Elasticsearch.
We refer to this as ``multi-field'' BM25 because it ingested the title and body of documents into separate fields (called ``title'' and ``contents'', respectively) in cases where the original corpus provided this information.
For corpora that didn't, all content was ingested into the default ``contents'' field.
Search was performed by generating a Lucene multi-field query that assigned both fields equal weight.
For corpora that did not explicitly have titles, the multi-field queries yielded the same ranking as if only the main ``contents'' field had been indexed and queried.

Building a baseline using Elasticsearch has the disadvantage in that it exists as an out-of-process retriever, which creates additional friction for researchers who desire a simple development/evaluation cycle.
This was discussed by~\citet{Devins_etal_WSDM2022}, who noted that since Elasticsearch was built on the open-source Lucene search library, researchers could ``bypass'' the features offered by Elasticsearch to directly gain in-process access to retrieval capabilities.
From the perspective of batch IR evaluations such as BEIR, this was expedient because the additional layers that Elasticsearch builds on top of Lucene provide little value to researchers.

A later iteration of the BEIR evaluation resources moved from Elasticsearch to the Pyserini IR toolkit, but this feature was never refined into a reproducible baseline that could be easily invoked by researchers.
In this work, we complete this ``packaging'' and explore additional BM25 variants (see Section~\ref{section:variants}).


\subsection{Learned Dense Retrieval Models}

We examine two learned dense retrieval models.
This class of models exhibits two key characteristics:\ use of dense semantic representations for retrieval and encoders for generating these representations that are trained with labeled datasets.

\paragraph{TAS-B} 
This is a BERT-based dense retrieval model proposed by~\citet{Hofstatter_etal_SIGIR2021}, where the primary innovation is a Balanced Topic Aware Sampling (TAS-B) strategy to assemble training batches for optimizing retrieval effectiveness in a data-efficient manner.
It was one of the earliest dense retrieval models to successfully exploit knowledge distillation, using dual supervision from a cross-encoder model and ColBERT.
TAS-B was one of the first dense retrieval models to be applied to BEIR, and was discussed in the original paper by~\citet{beir}.

\paragraph{Contriever} 
This is a dense retrieval model proposed by~\citet{Izacard:2112.09118:2021} that first applies retrieval-specific pretraining in an unsupervised manner (an Inverse Cloze Task variant) before fine-tuning with the MS MARCO passage dataset to optimize for retrieval effectiveness.
Contriever also builds on a BERT backbone and was specifically designed to explore zero-shot domain transfer capabilities.
It is among the most effective dense retrieval models available today on the BEIR benchmark.

\begin{table*}[t]
\centering
\begin{tabular}{lcccccccccc}
\toprule
\multirow{2}{*}{\textbf{Dataset}} & \multicolumn{5}{c}{\bf nDCG@10} & \multicolumn{5}{c}{\bf Recall@100} \\
 & BM25 & uniCOIL & SPLADE & TAS-B & Contriever & BM25 & uniCOIL & SPLADE & TAS-B & Contriever \\
 \cmidrule(lr){1-1} \cmidrule(lr){2-6} \cmidrule(lr){7-11}
TREC-COVID& 0.656& 0.640& 0.711& 0.505& 0.596& 0.114& 0.111& 0.131& 0.090& 0.091 \\
BioASQ& 0.465& 0.477& 0.504& 0.371& 0.383 & 0.715& 0.731& 0.742& 0.598& 0.607 \\
NFCorpus& 0.325& 0.333& 0.345& 0.324& 0.328& 0.250& 0.257& 0.289& 0.284& 0.301 \\
NQ& 0.329& 0.425& 0.544& 0.465& 0.498& 0.760& 0.833& 0.929& 0.904& 0.925 \\
HotpotQA& 0.603& 0.667& 0.686& 0.584& 0.638& 0.740& 0.798& 0.814& 0.728& 0.777\\
FiQA-2018& 0.236& 0.289& 0.351& 0.296& 0.329& 0.544& 0.553& 0.630& 0.582& 0.656 \\
Signal-1M& 0.330& 0.275& 0.296&	0.288& 0.278 & 0.370& 0.313& 0.331& 0.304& 0.322\\
TREC-NEWS& 0.398& 0.374& 0.394& 0.394& 0.428 & 0.422& 0.357& 0.432& 0.454& 0.492 \\
Robust04& 0.407& 0.403& 0.458& 0.461& 0.473 & 0.375& 0.317& 0.377& 0.411& 0.392 \\
ArguAna& 0.414& 0.396& 0.521& 0.436& 0.446& 0.943& 0.923& 0.982& 0.945& 0.977 \\
Tóuche-2020 {\small(v2)}& 0.367& 0.298& 0.243& 0.222& 0.204 & 0.538& 0.485& 0.472& 0.526& 0.442 \\
CQADupStack& 0.299& 0.301& 0.341& 0.309& 0.345 & 0.606& 0.569& 0.651& 0.612& 0.663\\
Quora& 0.789& 0.662& 0.814& 0.835& 0.865 & 0.973& 0.948& 0.982& 0.986& 0.994 \\
DBPedia& 0.313& 0.338& 0.442& 0.384& 0.413 & 0.398& 0.441& 0.564& 0.499& 0.541 \\
SCIDOCS& 0.158& 0.144& 0.159& 0.146& 0.165 & 0.356& 0.328& 0.367& 0.332& 0.378 \\
FEVER& 0.753& 0.812& 0.796& 0.733& 0.758 & 0.931& 0.955& 0.955& 0.945& 0.949 \\
Climate-FEVER& 0.213& 0.182& 0.228& 0.237& 0.237 & 0.436& 0.418& 0.514& 0.553& 0.575 \\
SciFact& 0.665& 0.686& 0.699& 0.644& 0.677 & 0.908& 0.912& 0.927& 0.894& 0.947 \\
\midrule
Avg.& 0.429& 0.428& 0.474& 0.424& 0.448 & 0.577& 0.569& 0.616& 0.591& 0.613 \\
\arrayrulecolor{black}
\bottomrule
\end{tabular}
\vspace{0.25cm}
\caption{Effectiveness results of five retrieval models across all 18 datasets in BEIR:\ nDCG@10 (left) and recall@100 (right).}
\label{table:main_results}
\end{table*}

\subsection{Learned Sparse Retrieval Models}

We examine two learned sparse retrieval models.
Like their dense counterparts, these models rely on an approach to retrieval based on representation learning that exploits labeled datasets.
However, these models generate sparse lexical representations instead of dense semantic ones.

\paragraph{uniCOIL (noexp)}
This model, originally proposed by~\citet{unicoil}, is a variant of COIL~\cite{gao-etal-2021-coil}, where BERT is trained to assign scalar weights to document tokens based on manually labeled relevance data (the MS MARCO passage dataset) to optimize retrieval effectiveness.
In the full setting, uniCOIL depends on a separate document expansion model~\cite{Ma_etal_SIGIR2022}, but here we use the ``no expansion'' (noexp) variant, which allows us to examine the domain transfer capabilities of a ``basic'' learned term weighting function.

\paragraph{SPLADE} 
This refers to a family of sparse retrieval models~\cite{spladev2} that learns both document/query expansion and term weighting with the help of a regularization factor to induce sparsity. 
As far as we are aware, this represents the state of the art in sparse retrieval models, particularly in a zero-shot setting.
More precisely, we use the SPLADE distil CoCondenser model,\footnote{Available for download here: \url{http://download-de.europe.naverlabs.com/Splade_Release_Jan22/splade_distil_CoCodenser_medium.tar.gz}} which as the name implies, uses distillation and the pretrained CoCondenser model~\cite{cocondenser}.

\section{Main Results}
\label{section:results}

The effectiveness of the five models presented in the previous section is shown in Table~\ref{table:main_results}, with nDCG@10 in the left group of columns and recall@100 in the right group of columns.
Each row corresponds to one of the BEIR datasets, and the rows are ordered in the same manner as \citet{beir}.

\begin{figure*}[t]
\centering
\includegraphics[width=0.47\textwidth]{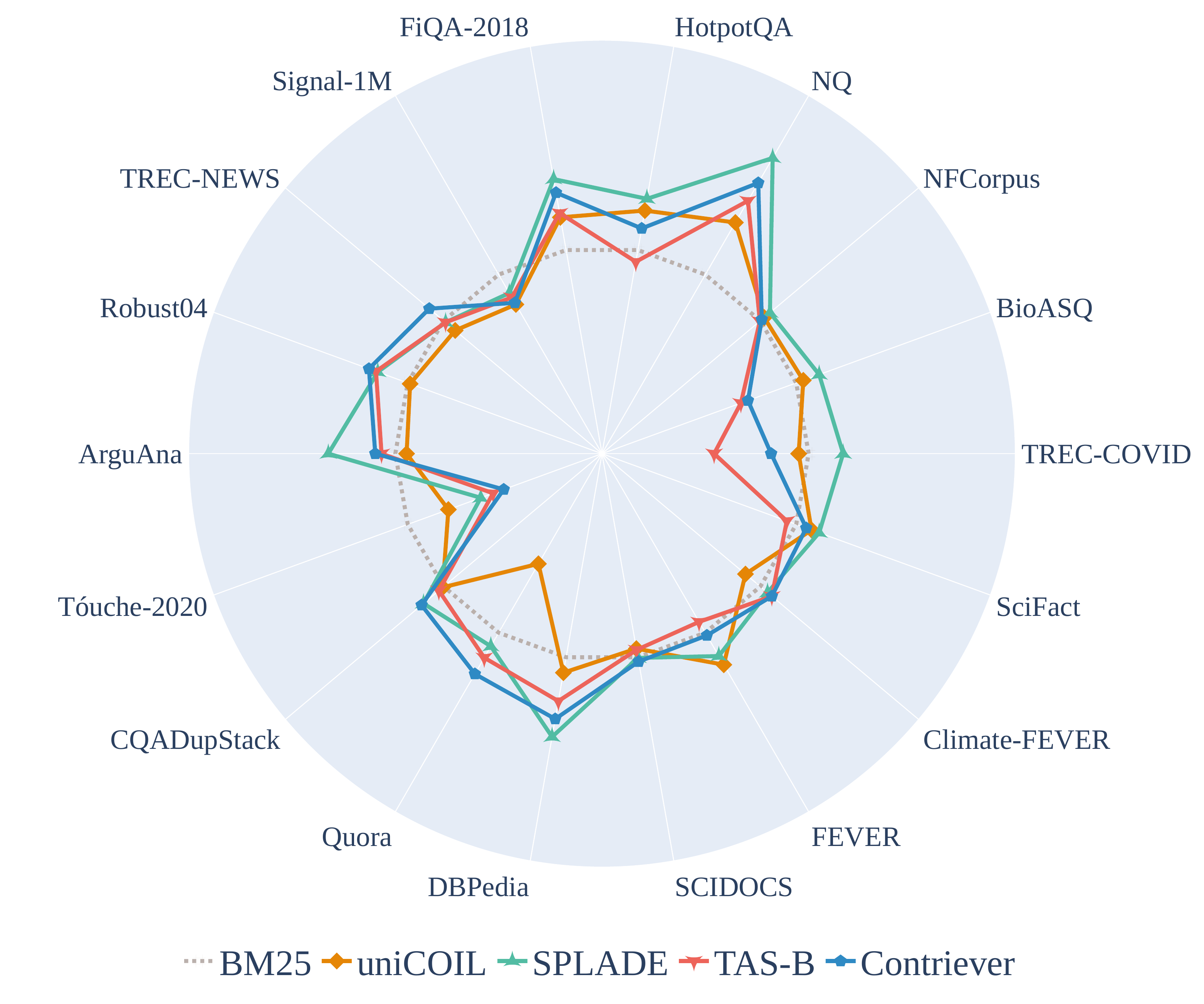}
\includegraphics[width=0.47\textwidth]{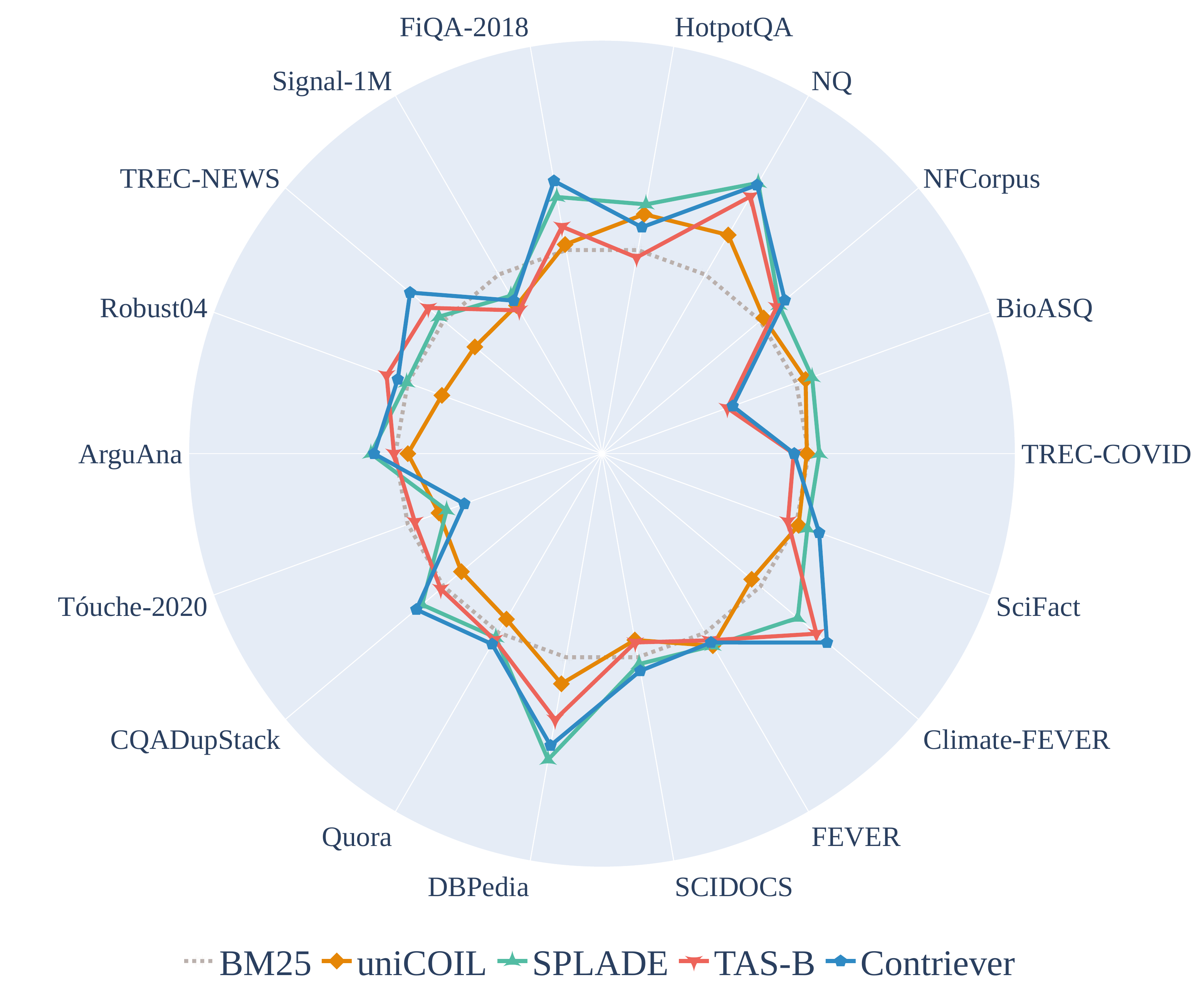}
\caption{Radar charts comparing the effectiveness of five retrieval models across all 18 datasets in BEIR:\ nDCG@10 (left) and recall@100 (right). The effectiveness of BM25 (dotted line) is scaled to half of the radius of the chart area, and the effectiveness of the other models is scaled accordingly.}
\label{fig:beir-radar}
\end{figure*}

We have seen previous papers report BEIR results using slightly different layouts and organizations, which make comparisons difficult.
Moving forward, we offer a few best practices to promote consistency in how results are shared:
We feel that presenting the datasets in rows and effectiveness metrics in columns feels more natural, and urge the community to also adopt this layout.
The alternative of showing the different datasets in columns feels more awkward to us.
Furthermore, we recommend that researchers order the rows exactly as \citet{beir}, which we have done here.
Other reasonable alternatives, for example, alphabetical sorting, discard the ``semantic grouping'' of the datasets.

Encouraging consistency in the presentation of BEIR results is an important first step to gaining insight when comparing retrieval models.
However, there is no hiding the fact that BEIR scores comprise a complex aggregation of diverse datasets, and the standard approach of comparing macro-averaged nDCG@10 scores (as we have done in the final row of Table~\ref{table:main_results}) is deficient in many ways.

It is well known that averages often hide important individual differences, but teasing apart these differences from a large table of numbers such as Table~\ref{table:main_results} can be difficult.
For example, the results show that uniCOIL and BM25 achieve a similar level of effectiveness overall (0.428 vs.\ 0.429), but what can we say about effectiveness on individual datasets?
Glancing down the rows, we see many differences---some large, some small---so it is possible to conclude that although uniCOIL and BM25 are ``about the same'' averaged across the BEIR datasets, effectiveness on individual datasets differ.
How can we gain more insight easily?
The same question applies when comparing BM25 and TAS-B, where the average nDCG@10 scores are comparable.
Consider the SPLADE and Contriever results:\ we see that both achieve a higher average across all the datasets, but is this due to consistent gains across many datasets or a few big gains?
It's difficult to tell from Table~\ref{table:main_results}.

We present a potential solution to these challenges in terms of radar charts:\
Figure~\ref{fig:beir-radar} shows a visualization comparing the effectiveness of the five retrieval models, with nDCG@10 on the left and recall@100 on the right.
Each radar chart comprises 18 axes, arranged radially, in the same order as the rows in Table~\ref{table:main_results}.
The effectiveness of a model is plotted on each of the 18 axes and connected by line segments to form a polygon.
The effectiveness of BM25, which serves as a baseline, is scaled to half of the radius of the entire chart area, so the effectiveness of BM25 across all 18 datasets is captured by the dotted polygon.
The effectiveness of the other models on each corpus is scaled relative to BM25.
That is, points further away from the center represent higher scores and points closer to the center represent lower scores, where the distance to the midpoint of the axis is proportional to the score difference with respect to BM25.

The radar charts allow us to easily compare the effectiveness of the models across all datasets, and differences that are obscured by averages come readily to light.
Focusing on nDCG@10, consider the question above about BM25 vs.\ uniCOIL (orange):\
We can see that uniCOIL excels on HotpotQA and NQ in terms of nDCG@10, but otherwise achieves effectiveness that is either on par with BM25 or worse.
In particular, on Signal-1M, Quora, and T\'ouche-2020, uniCOIL is substantially worse.
We see a similar situation with TAS-B, which is more effective on some datasets but performs terribly on others, most notably TREC-COVID, BioASQ, and T\'ouche-2020.
Inconsistent effectiveness is similarly observed with Contriever as well, even though on average the model scores higher than BM25.
It appears that both dense retrieval models perform rather poorly on BioASQ and TREC-COVID, two datasets that focus on biomedical retrieval.
For all the models examined, it appears that SPLADE exhibits the most consistent gains, with only a few datasets where nDCG@10 is worse than BM25.

Focusing on recall@100 (Figure~\ref{fig:beir-radar}, right), we believe that the radar chart is similarly helpful in highlighting effectiveness differences between the models that are obscured by only looking at the means.
In particular, conclusions drawn from the nDCG@10 scores differ from those based on recall@100:\ both metrics are important, but for different reasons.
Early precision metrics such as nDCG@10 directly quantify output quality if results from a first-stage retriever are directly presented to the user.
On the other hand, recall@100 quantifies the upper bound of reranking effectiveness.
Based on the radar chart, we clearly see that neural models do {\it not} consistently increase effectiveness.
We observe large gains---for example, all four models perform very well on NQ and DBPedia---but we also observe many cases where at least some of the neural models perform poorly---for example, on Signal-1M and BioASQ.

We have some explanations for these findings.
First, let us consider the poor effectiveness of dense retrieval models on TREC-COVID and BioASQ.
BioASQ involves scientific paper retrieval given a biomedical query that often involves specialized terminology (and similarly for TREC-COVID as well).
As an example, consider the BioASQ query, ``Is AZD5153 active in prostate cancer?''
Here, ``AZD5153'' is a specialized biomedical term, which the model most likely has seen only rarely during training.
This would cause issues for dense retrievers, as they are unable to represent rare terms well within the embedding space, and hence retrieval effectiveness would suffer.
That is, poor effectiveness can be explained by domain shifts between training data (general web) and the test data (biomedical texts).
On the other hand, sparse models such as SPLADE or uniCOIL produce representations that retain lexical matching, and thus are less impacted by domain shifts involving vocabulary differences.
The relatively poor effectiveness of models on Signal-1M (tweets) can be explained similarly.

Second, we observe that all transformer models achieve big gains on NQ.
Previous work has found that effectiveness on NQ and effectiveness on the TREC 2019 Deep Learning (DL) track have the highest correlation among all BEIR datasets in terms of scores \cite{10.1145/3511808.3557312} (that is, models that perform well on TREC DL also perform well on NQ).
The close connections between the MS MARCO datasets and the TREC Deep Learning tracks suggest that transfer from MS MARCO is particularly advantageous for the NQ dataset.
Since all models examined in this paper take advantage of MS MARCO, the big gains on NQ might be a data artifact and not a demonstration of robust domain transfer capabilities.

Finally, it appears that all transformer models perform worse than the BM25 baseline for T\'ouche-2020, sometimes by sizeable margins.
Deeper analysis suggests a number of factors at play.
The dataset has the fewest number of queries (49) and the relevance judgments were created from very shallow (top 5) pools.
As a point of contrast, TREC-COVID, the dataset with the next smallest number of queries (50), has much deeper pools.
T\'ouche-2020 queries are rather short, for example, ``Is obesity a disease?''\ and phrased in a manner where relevant documents are highly likely to contain query terms.
The task of argument retrieval is quite different from standard {\it ad hoc} retrieval tasks, thus potentially hurting zero-shot transfer effectiveness.
Reading through the organizers' description of the evaluation, it appears that most submissions focused on lexical techniques, and we believe that the relevance judgments are biased towards models based on lexical matches.
Thus, the qrels might be unable to fairly assess models that retrieve semantically relevant content but do not necessarily contain query terms.
We believe this is the case for many of the datasets in BEIR, although this phenomenon is most pronounced in T\'ouche-2020.


Our point here is not to exhaustively explain all the effectiveness differences observed, although we do offer some analyses above.
Instead, the main contribution of this paper is a methodological tool (i.e., visualizations using radar charts) that provides a starting point for further analyses.
For researchers, it would be worthwhile to follow up any of the findings above with more detailed error analyses, but such a detailed study is beyond the scope of this work.
Beyond support for error analyses, we believe that these radar chart visualizations are helpful for practitioners who might be interested in deploying neural models.
From the perspective of real-world applications, it would make sense to ensure that a deployed model ``performs no worse'' than BM25, and thus these results lead us to conclude that none of the models are viable as a replacement yet, given that there are clearly situations where effectiveness is substantially lower.

\section{Reproducible Implementations}

We provide reproducible implementations of the retrieval models discussed in this paper.
At a high level, our goal is to make it as easy as possible for researchers to reproduce the results in Table~\ref{table:main_results}.
To be precise, here we are using reproducibility in the sense articulated by the ACM in its Artifact Review and Badging Policy,\footnote{\url{https://www.acm.org/publications/policies/artifact-review-and-badging-current}} characterized as ``different team, same experimental setup''.
Specifically, ``this means that an independent group can obtain the same result using the author's own artifacts.''

Our reproducible implementations conform to the aspirational ideal of ``two-click reproductions'' described by~\citet{Lin_arXiv2022}.
The motivation is that a user should be able to reproduce an experimental result with only two clicks:\ a ``copy'' and a ``paste'' from a documentation page.
That is, the user will arrive at the same nDCG@10 and recall@100 reported in Table~\ref{table:main_results}.

Our reproducible implementations leverage previous efforts and infrastructure investments in the Pyserini IR toolkit~\cite{Lin_etal_SIGIR2021_Pyserini}, which is built on Anserini~\cite{Yang_etal_JDIQ2018}.
Anserini is an IR toolkit built on the open-source Lucene search library, and like Lucene, it was written in Java.
To provide compatibility with Python, the dominant language for building neural retrieval models today, we developed Pyserini, which provides Python bindings for Anserini as well as many other non-Java capabilities.

In our design, BM25 baselines and sparse retrieval models are directly implemented in Anserini with Lucene inverted indexes, exposed in Pyserini in Python.
The dense retrieval models are implemented using the 
 Faiss library for efficient similarity search and clustering of dense vectors~\cite{Johnson_etal_2021} by Meta Research; for simplicity, we used flat indexes.
Pyserini provides a uniform API to support retrieval using all the models, for example, abstracting over the Java-based implementation of retrieval using BM25 and the sparse retrieval models.

To provide a concrete example, performing a BM25 retrieval run over the test queries in the BioASQ corpus in BEIR can be accomplished by the following command:

\begin{small}
\begin{verbatim}
    python -m pyserini.search.lucene \
      --index beir-v1.0.0-bioasq-multifield \
      --topics beir-v1.0.0-bioasq-test \
      --output run.beir-multifield.bioasq.txt \
      --output-format trec \
      --batch 36 --threads 12 \
      --hits 1000 --bm25 --fields contents=1.0 title=1.0   
\end{verbatim}
\end{small}

\noindent The main driver program for searching Lucene inverted indexes is \mytt{pyserini.search.lucene}.
In this example, we are using multi-field BM25, with equal weights set to both the ``title'' and ``contents'' fields (by default), specified using the \mytt{{-}{-}fields} command-line argument.
The remaining arguments are mostly self-explanatory, but we provide additional commentary:

The \mytt{{-}{-}index} argument specifies a pre-built inverted index for the BioASQ corpus that is stored on our group's servers.
On the first invocation of the above command, the driver automatically downloads the index and caches it on the local machine.
The \mytt{{-}{-}topics} argument specifies the BioASQ test queries, which are already included as part of Pyserini.
With this design, the user does not need to separately figure out where to download the indexes and queries to successfully reproduce a result.

Pyserini also includes all the components necessary to evaluate the retrieval results.
In this case, the nDCG@10 score can be computed as follows:

\begin{small}
\begin{verbatim}
    python -m pyserini.eval.trec_eval \
      -c -m ndcg_cut.10 beir-v1.0.0-bioasq-test \
      run.beir-multifield.bioasq.txt
\end{verbatim}
\end{small}

\noindent We provide a wrapper around the \mytt{trec\_eval} package, and relevance judgments are included in Pyserini.
Once again, this saves the user additional effort in needing to track down evaluation tools and relevance judgments from various web sources.

Modern IR evaluation methodology can be quite complex, but with our ``two click reproductions'', the two commands above will produce the results in Table~\ref{table:main_results}.
We have built a landing page\footnote{\url{https://castorini.github.io/pyserini/2cr/beir.html}} in the Pyserini documentation that provides an entry point to a ``reproduction matrix'' comprising all models and all datasets.

\section{Leaderboard}
\label{section:leaderboard}

Another major contribution of this work is the introduction of an official self-service BEIR leaderboard that unifies community-wide evaluation efforts.
This replaces an informal leaderboard that was initially maintained via a public Google spreadsheet with read-only permissions.
Updating the leaderboard involved researchers directly contacting the organizers to share the effectiveness of each task, who would manually enter the new results into the spreadsheet.
In the beginning, this was manageable because there were relatively few researchers, but after BEIR started to gain traction in the community, the overhead of communications and manual effort made leaderboard maintenance unsustainable.
As a result, the spreadsheet leaderboard became stale as time went by.
And as the pace of leaderboard updates slowed, researchers sought other means to disseminate results.
This in turn led to disparate forms of reporting results.
Many researchers opted to report their results on arbitrary subsets of BEIR, giving rise to scores on several different subsets containing anywhere from 11 to 15 datasets.

\begin{table*}[t]
\centering
\begin{tabular}{lccccccccc}
\toprule
\multirow{2}{*}{\textbf{Task}}  & \multicolumn{3}{c}{\textbf{BM25}} & \multicolumn{3}{c}{\textbf{TASB}} & \multicolumn{3}{c}{\textbf{Dense--Sparse Hybrid}}\\
 & multifield & flat & flat-wp & FirstP & MaxP (10/5) & MaxP (8/4) & Contriever & SPLADE & Hybrid\\
 \cmidrule(lr){1-1} \cmidrule(lr){2-4} \cmidrule(lr){5-7} \cmidrule(lr){8-10}
\textbf{TREC-COVID}& 0.656& 0.595& 0.565& 0.481& 0.491& 0.505 & 0.596 & 0.711 & 0.730 \\
\textbf{BioASQ}& 0.465& 0.523& 0.419& 0.360& 0.367& 0.371 & 0.383 & 0.504 & 0.478 \\
\textbf{NFCorpus}& 0.325& 0.322& 0.314& 0.319& 0.321& 0.324 & 0.328 & 0.345 & 0.352 \\
\textbf{NQ}& 0.329& 0.306& 0.305& 0.463& 0.463& 0.465 & 0.498 & 0.544 & 0.562 \\
\textbf{HotpotQA}& 0.603& 0.633& 0.593& 0.584& 0.584& 0.584 & 0.638 & 0.686 & 0.700 \\
\textbf{FiQA-2018}& 0.236& 0.236& 0.218& 0.300& 0.295& 0.296 & 0.329 & 0.351 & 0.374 \\
\textbf{Signal-1M}& 0.330& 0.330& 0.350& 0.288& 0.288& 0.289 & 0.278 & 0.296 & 0.302 \\
\textbf{TREC-NEWS}& 0.398& 0.395& 0.361& 0.377& 0.398& 0.394 & 0.428 & 0.394 & 0.458 \\
\textbf{Robust04}& 0.407& 0.407& 0.377& 0.428& 0.455& 0.461 & 0.473  & 0.458 & 0.505 \\
\textbf{ArguAna}& 0.414& 0.397& 0.364& 0.427& 0.433& 0.436 & 0.446 & 0.521 & 0.520 \\
\textbf{Tóuche-2020 {\small(v2)}}& 0.367& 0.442& 0.466& 0.163& 0.215& 0.222 & 0.204 & 0.243 & 0.226 \\
\textbf{CQADupStack}& 0.299& 0.302& 0.295& 0.314& 0.309& 0.309 & 0.345 & 0.341 & 0.369 \\
\textbf{Quora}& 0.789& 0.789& 0.730& 0.835& 0.835& 0.835 & 0.865 & 0.814 & 0.862 \\
\textbf{DBPedia}& 0.313& 0.318& 0.284& 0.384& 0.384& 0.384 & 0.413 & 0.442 & 0.455 \\
\textbf{SCIDOCS}& 0.158& 0.149& 0.138& 0.149& 0.147& 0.146 & 0.165 & 0.159 & 0.176 \\
\textbf{FEVER}& 0.753& 0.651& 0.658& 0.700& 0.724& 0.733 & 0.758 & 0.796 & 0.814 \\
\textbf{Climate-FEVER}& 0.213& 0.165& 0.158& 0.228& 0.241& 0.237 & 0.237 & 0.228 & 0.265 \\
\textbf{SciFact}& 0.665& 0.679& 0.672& 0.643& 0.645& 0.644 & 0.677 & 0.699 & 0.734 \\
\cmidrule(lr){1-1} \cmidrule(lr){2-4} \cmidrule(lr){5-7} \cmidrule(lr){8-10}
Avg.nDCG@10& 0.429& 0.424& 0.404& 0.414& 0.422& 0.424 & 0.448 & 0.474 & 0.493 \\
\arrayrulecolor{black}
\bottomrule
\end{tabular}
\vspace{0.25cm}
\caption{Effectiveness (in terms of nDCG@10) of model variants across all 18 datasets in BEIR.}
\label{table:variants}
\end{table*}

As a remedy to these issues, we have built an official leaderboard to standardize evaluation on BEIR, which is linked from the website at \url{http://beir.ai/}.
The new leaderboard is hosted on EvalAI \cite{evalai}, which is an open-source platform designed to enable human-in-the-loop and interactive agent evaluations.
It has successfully hosted over 200 challenges to date from more than 30 organizers.
The platform provides the capability to configure customized evaluation pipelines with flexible dataset splits and challenge phases that can be executed on cloud-based worker pools.
Specifically, EvalAI defines a new challenge through a competition bundle, encapsulated in a GitHub repository, that contains the challenge configuration (e.g., start/finish date, submission limits, etc.), evaluation code, and information about data splits.
Moreover, it supports the processing of large submission files, making it particularly suitable for the BEIR benchmark, which often involves large run files. 

The new leaderboard offers an easy-to-use interface accompanied by documentation, providing a step-by-step description of the submission protocol. 
To evaluate a model on BEIR, users need to sign up for EvalAI first. 
Then, they can submit their run files through the EvalAI command-line library or through a web interface. 
Each run file may contain at least top-10 and up to top-100 retrieved results.
Upon receiving a submission, an evaluation script validates each run file, promptly addressing any edge cases that may arise.
For example, queries that are returned in retrieved documents will be discarded.
The script subsequently measures the effectiveness of each task.
The results can be readily viewed in the public leaderboard, although users have the option to keep their results private.
We limit the submission rate to one per day.

Moving forward, this leaderboard will serve as the central ``source of truth'' for BEIR results, enforcing consistency in evaluation protocol.
Furthermore, we have invited the community to submit results from their already published methods as an attempt to consolidate results that pre-date the leaderboard.

\section{Model Variants}
\label{section:variants}

The models presented in Section~\ref{section:models} cover the major approaches to neural retrieval today.
In this section, we further examine variants that help us better understand some of the strengths and limitations of those models.

\subsection{Multi-Field Indexing}

A baseline ``as simple as BM25'' still presents a number of design decisions that may impact effectiveness in substantive ways.
One such choice made by~\citet{beir} in the initial BEIR release is the use of multi-field indexing in the BM25 baseline.
That is, the title and main body of each document is separately indexed, inserted into the ``title'' and ``contents'' fields, respectively.
At search time, a multi-field (Lucene) query combines evidence from both fields (with equal weights).

\begin{figure*}[t]
\centering
\includegraphics[width=0.33\textwidth]{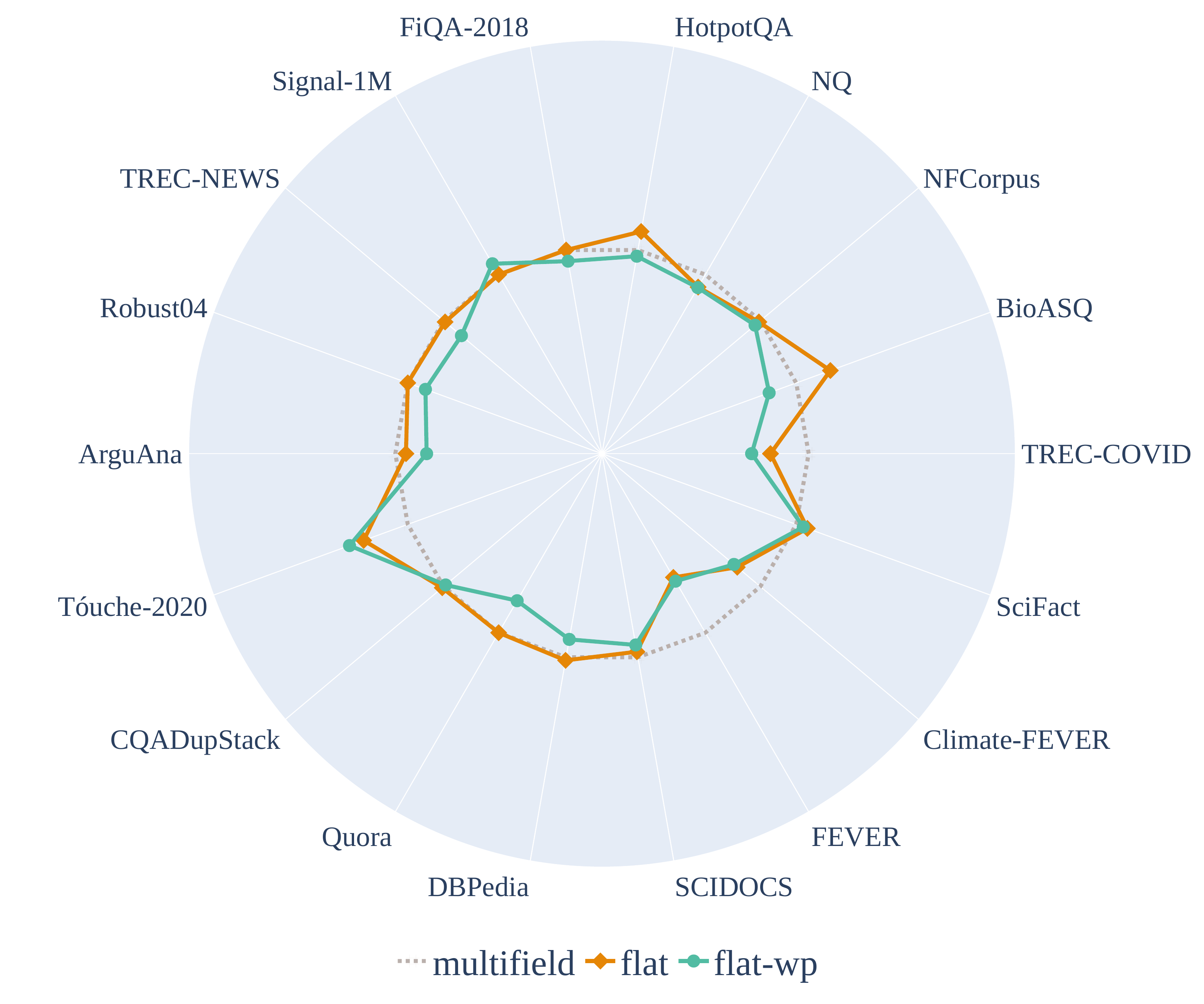}
\includegraphics[width=0.33\textwidth]{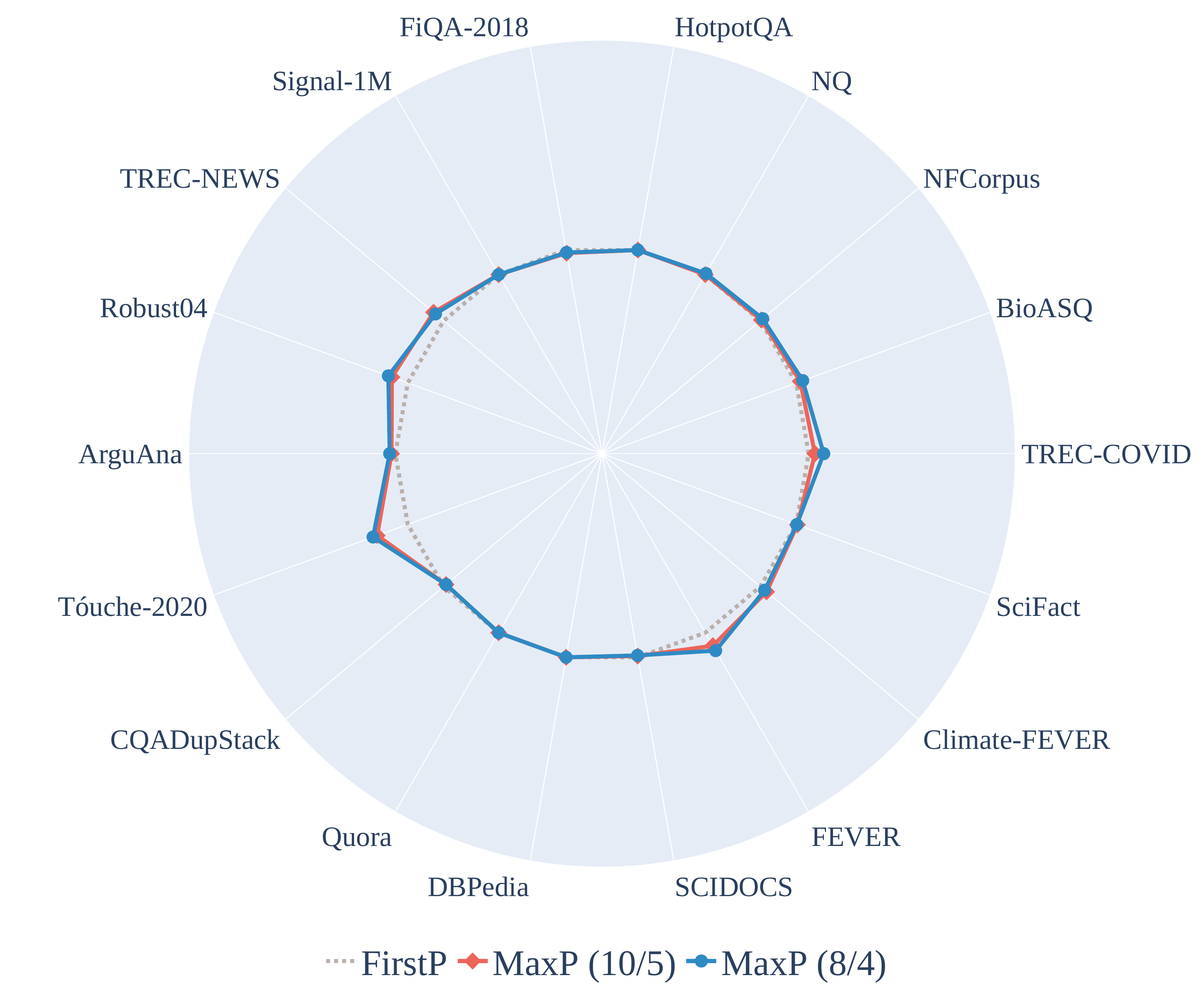}
\includegraphics[width=0.33\textwidth]{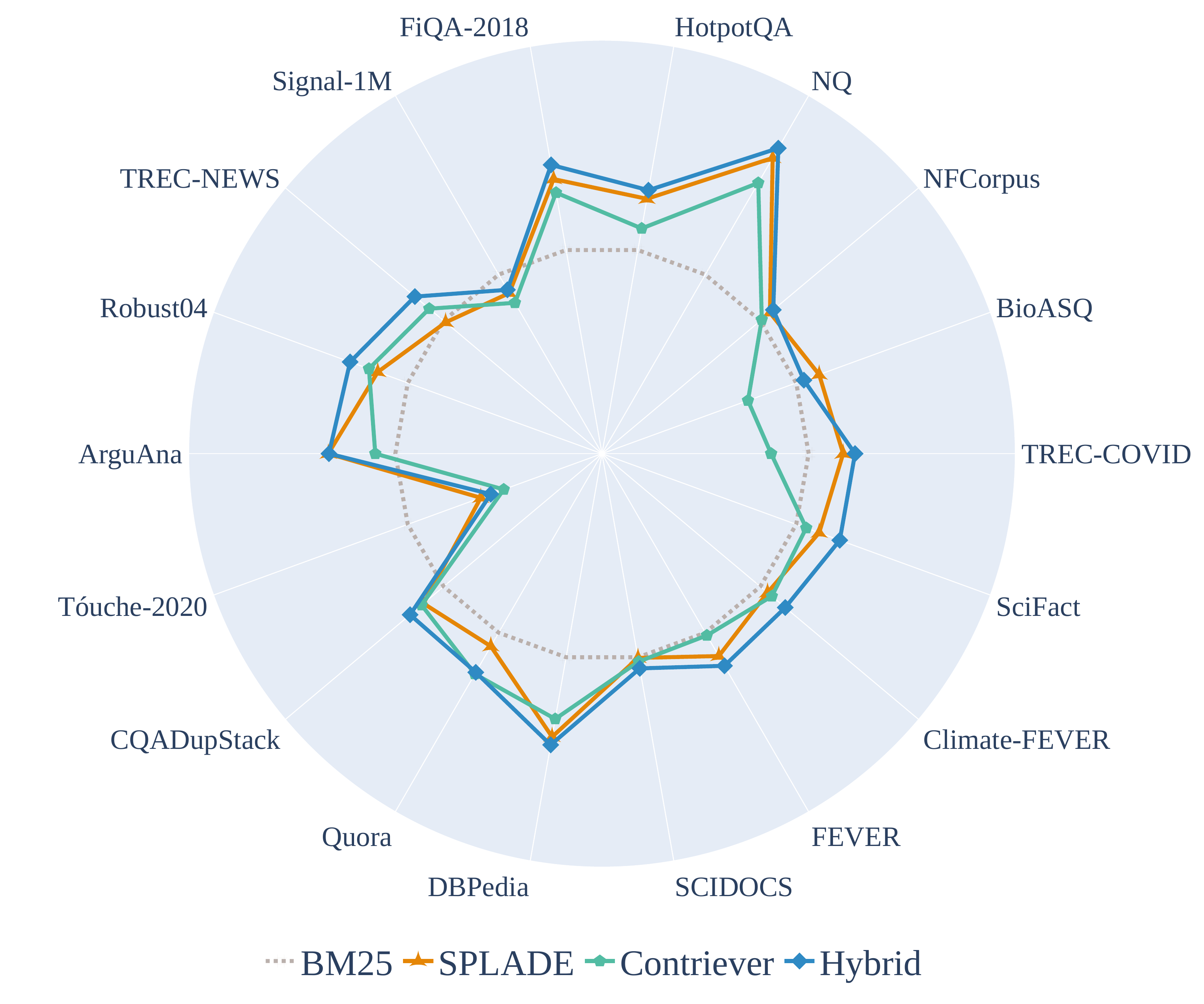}
\caption{Radar charts visualizing the effectiveness (in terms of nDCG@10) of different model variants:\ BM25 variants (left), techniques for searching long documents (middle), and dense--sparse hybrids (right).}
\label{fig:radar-variants}
\end{figure*}

What is the impact of this document structure on effectiveness?
We can answer this question with a variant that we call ``flat'' BM25, where the title and main body of each document are concatenated together and indexed in a single field.
These results are shown in Table~\ref{table:variants} under the ``flat'' column; the ``multifield'' column refers to the default BM25 configuration from Table~\ref{table:main_results}.
Following the analyses in Section~\ref{section:results}, the radar chart visualization comparing ``flat'' to ``multifield'' BM25 is shown in Figure~\ref{fig:radar-variants}, with the latter configuration as the reference.
The radar chart shows that in some cases ``flat'' is better (e.g., BioASQ, HotpotQA, and Tóuche-2020) and in other cases, it is worse (e.g., TREC-COVID, FEVER, and Climate-FEVER), but overall the differences are relatively small.
It is hard to draw reliable conclusions, as these differences primarily stem from corpus organization, which obviously varies across the datasets.

Another important decision in building a BM25 baseline is the choice of tokenization and stemming.
In Lucene, an abstraction called the analyzer is responsible for converting a sequence of bytes into a sequence of tokens.
In all the BM25 variants discussed above, we used Lucene's default analyzer for English.
In contrast, the sparse representation models use BERT's wordpiece vocabulary.
Since the two vocabulary spaces are different, one might argue that comparisons are not fair.
To examine these effects, we applied the wordpiece tokenizer to the ``flat'' BM25 condition.
The results are shown in the ``flat-wp'' column of Table~\ref{table:main_results}, and visualized in the radar chart shown in Figure~\ref{fig:radar-variants} (left).
Once again, the differences are relatively small, but it does appear that wordpiece tokenization consistently degrades effectiveness.
This occurs because wordpiece tokenization often chops long content words into shorter subwords that are polysemous, hence introducing noise.

\subsection{Searching Long Documents}

One well-known issue with retrieval methods built on pretrained transformers is that the underlying models have length restrictions in input text; see~\citet{Lin_etal_2021_ptr4tr} for extensive discussions of this topic.
The two commonly adopted solutions are to either encode only the first $N$ tokens in each document or to segment a longer document into passages and encode each passage independently.
In the terminology of~\citet{Dai_Callan_SIGIR2019}, these approaches are known as FirstP and MaxP, respectively.
With MaxP, multiple representations are generated per document, and at retrieval time, the maximum of the passage scores is taken as the score of the document; this heuristic itself dates back to at least the 1990s~\cite{Hearst_SIGIR1993,Callan_SIGIR1994}.

Curiously, most papers that report evaluations on BEIR contain no explicit discussions about how long documents were processed.
Based on informal communications with model developers and examination of available open-source implementations, it appears that most researchers apply the FirstP approach.
That is, they simply truncate each document to the first $N$ tokens (where $N$ varies by model).
This, of course, begs the question of whether different techniques for searching long documents ``make a difference''.

We conducted experiments to answer this question.
Given the vast design space of options for segmenting longer documents into shorter passages, we built on previous work that explored some of the design choices.
Following~\citet{Pradeep_etal_arXiv2021_EMD} and later work by~\citet{Ma_etal_SIGIR2022}, we decided to segment documents into sliding windows of sentences.
Based on their previous explorations, we examined two configurations:\ a sliding window of 10 sentences with a stride of 5 sentences, and an 8/4 combination.
Passages based on sentences yield variable-length passages, in contrast to the obvious alternative of using fixed-length windows.
However, sentence-based windows preserve natural discourse units and better encapsulate context that might be useful for determining relevance.
For these experiments, we used the sentence chunker in spaCy (version 3.4.4).

Experimental results are presented in Table~\ref{table:variants}, shown for the dense retrieval model TAS-B; the corresponding radar chart is shown in Figure~\ref{fig:radar-variants} (middle).
The visualization makes it clear that MaxP does yield some gains---in 15 out of the 18 datasets---but the overall differences are small.
Based on these results, we would argue that to evaluate future models, FirstP is ``good enough'', since methodological consistency is likely more important.
That is, comparing FirstP on dense model $A$ with MaxP on dense model $B$ would introduce confusion and conflate unrelated factors.

\subsection{Hybrid Fusion}

One clear takeaway from Section~\ref{section:results} is that the effectiveness of zero-shot transfer for both learned dense and learned sparse representations is inconsistent across the 18 BEIR datasets.
From the radar charts, we see some clear gains, for example, on NQ, but also cases where some models underperform, most notably, dense retrieval models on BioASQ.
In such cases, hybrid fusion techniques can perhaps be helpful in combining evidence from different sources.
While this general idea dates back several decades at least~\cite{Bartell_etal_SIGIR1994}, more recent work has demonstrated that fusion between lexical and semantic representations work particularly well~\cite{Gao_etal_ECIR2021,Ma_etal_arXiv2021_DPR}.
Furthermore, the fusion of BM25 with dense~\cite{xu2022laprador} and sparse~\cite{spladev2} representations has already been shown to be effective.

Here, we explored fusion techniques further, primarily to see whether we can combine multiple sources of evidence to achieve {\it consistent} gains across all BEIR datasets.
We applied the simple dense--sparse hybrid fusion techniques described by~\citet{Ma_etal_arXiv2021_DPR} to combine Contriever and SPLADE, the most effective dense and sparse retrieval models, respectively.
Specifically, we first retrieved the top 1000 documents separately using each model.
We then normalized the relevance scores from each source into the range $[0, 1]$ and computed the final hybrid score as the average of the two scores to produce new rankings for evaluation.

Experimental results are presented in Table~\ref{table:variants} in the rightmost column, ``Hybrid''.
The nDCG@10 figures for the two sources, Contriever and SPLADE, are copied from Table~\ref{table:main_results} for convenience.
We see that, in general, the hybrid approach is able to improve over the best individual model, with four exceptions:\
BioASQ, T\'ouche-2020, ArguAna, and Quora, although for the last two, the differences are quite small.
How does this fusion run compare to BM25?
The radar chart visualization that answers this question is shown in Figure~\ref{fig:radar-variants} (right), where we plot the effectiveness of Contriever, SPLADE, and our hybrid approach with BM25 as the reference.
Table~\ref{table:variants} shows more than a six-point gain on average, but more importantly, the radar chart visualization shows that the gains are consistent.
We see that the hybrid beats BM25 on all but two datasets:\ Signal-1M and T\'ouche-2020.
In particular, the visualization makes it clear that SPLADE is able to compensate for the poor effectiveness of Contriever on BioASQ and TREC-COVID, and in cases where Contriever is more effective than SPLADE, the hybrid approach further boosts effectiveness.
Simple score averaging seems to achieve the best of both worlds, and this dense--sparse hybrid appears to attain a level of robustness that none of the other models exhibit.

\section{Conclusions and Future Work}

The BEIR benchmark provides an important instrument for evaluating the cross-domain robustness of retrieval models and has gained traction due to the growing recognition of retrieval as a form of representation learning.
The efforts described in this paper address the two shortcomings that we have identified with BEIR:\ challenges in reproducibility and in the sharing of results.
Reproducible reference implementations in the Pyserini IR toolkit tackle the first challenge.
An official self-service leaderboard and best practices for sharing results target the second challenge.

Looking ahead, we see two more challenges that need to be addressed:
First, how do we perform significance testing for a benchmark like BEIR that is itself an aggregation of individual datasets?
There is no agreed-upon methodology, and although one could perform tests on each dataset, significance testing across the final macro-average would not be meaningful since the datasets diverge so much in corpus size, number of queries, number of judgments, and many other dimensions.
Thus, there is currently no way to answer the question:\ Is this retrieval model significantly better than that across a broad range of domains?
While our radar charts can show if observed gains are consistent across the BEIR dataset, they currently lack statistical rigor.

The second major issue concerns potential systematic biases in the current relevance judgments used in the evaluations.
Many of the datasets were created before the advent of modern transformer-based models, and may have relevance judgments biased towards lexical matching techniques.
\citet{Voorhees_etal_arXiv2022} argued that for old TREC collections that used deep pools, this doesn't appear to be an issue, but many of the datasets in BEIR used rather shallow pools (e.g., T\'ouche-2020), where potential biases are still poorly understood.
Operationally, these issues manifest as unjudged documents, sometimes called ``holes''.
We are not aware of any work that has examined these issues in detail, but we believe that such studies are sorely needed.

To end on a positive note, we are optimistic about the future of the BEIR benchmark.
It is an important evaluation instrument for the community to address a number of timely research questions.
This work mitigates two existing shortcomings, and while there remain more challenges ahead, BEIR has already and will continue to help advance the state of the art.

\section*{Acknowledgements}

This research was supported in part by the Canada First Research Excellence Fund and the Natural Sciences and Engineering Research Council (NSERC) of Canada.


\bibliographystyle{ACM-Reference-Format}
\bibliography{ref}

\end{document}